\documentclass[12pt]{article}  
\usepackage{epsfig} 
\usepackage{graphics} 
\usepackage{rotate} 

\def \bg{\bigskip} 
\def \no{\noindent} 
\begin{document}

 \normalsize

\setcounter{page}{1}

  rel1,   Oct.,  13,  1999  


\bg

\begin{center} 
{\large {\bf Radiation from a Charge in a Gravitational Field }}   

\end{center}

{\centerline  {\bf Amos Harpaz  \& Noam Soker }} 

\bg

{\centerline { Department of Physics, University of Haifa at 
Oranim, Tivon 36006, ISRAEL }}
   
 \no  phr89ah@vmsa.technion.ac.il 

 \no soker@physics.technion.ac.il

\bg
\no {\bf ABASTRACT} 

When an electric charge is 
supported at rest in a static gravitational field, its electric field is not 
supported with the charge, and it falls freely in the gravitational 
field.  Drawing the electric field lines continuously in time, we find 
 that they always emerge from the charge, but the 
 electric field is curved and there is a stress force 
between the freely  falling (curved) field and the static charge.  The 
 charge radiates and the   
work done by the gravitational field to overcome the stress force 
is the source for the energy  radiated by the supported (static) 
charge.  {\it  A static charge in a gravitational field radiates, 
as predicted by the principle of equivalence}.  
 This mechanism is similar to the one applied to an electric charge 
 accelerated in a free space.  In this case, the electric 
field is not accelerated with the charge.  The electric field is curved, 
and there is a stress force between the charge and its field.  The 
work done in overcoming the stress force is the source of the 
energy radiated by the accelerated charge.  
         
\bg

\no key words: Principle of Equivalence, Curved Electric Field 

\vfil

\eject

\bg

\no {\bf 1.  Introduction} 
\bg

The validity of the principle of equivalence (POE) to the case 
 of radiation from  
an electric charge in a gravitational field (GF) is a long-standing 
problem (refs. [1], [2], and references cited therein).  
 Specifically it is discussed in connection with
two cases: (1) Does an electric charge, freely falling in a gravitational 
field radiate?  (2) Does a charge supported at rest in a gravitational 
field radiate? 
  Using plainly the POE one may conclude that a freely falling charge in 
  a GF will not radiate because its situation is equivalent 
  to that of a free charge in empty space, and a charge supported at rest 
  in a GF (chracterized by an acceleration, $ g$), will radiate 
because its situation is equivalent to that of a charge accelerated in free 
space with an acceleration $g$.   The common approach in the physical society 
is the opposite one - it is believed that a static observer in a gravitational 
field will find that a freely falling charge in a GF  
 does  radiate, while a charge supported at rest in a GF  
 does not radiate [1].   It is also concluded that the validity of 
 the POE is limited, and it 
is not a general principle.  

However, this approach led to several 
contradictions, which in turn, led people to conclude that the ability to 
observe a radiation depends on the relative accelration between the 
charge and the observer:  an observer falling freely in parallel to a freely 
falling charge will not observe radiation, while a static observer in 
the same field will observe radiation.  In 
the same way, a static observer located in a lab where a charge 
  is supported at  rest in a GF will not    
  observe radiation, while a freely falling observer, 
 passing by the same charge, will observe radiation (ref [1], pp 218). 

The electromagnetic radiation is  defined as a relative phenomenon,   
that depends on the relative acceleration between the observer and the charge. 
In the following, we analyze the process that leads to 
  the creation of radiation. 
We demand that radiation as a process of energy transfer is a physical 
 event (which is an objective 
phenomenon), and we come to the conclusion that a freely falling charge 
 does not radiate, and a charge supported at rest in a GF does radiate. 
These conclusions are in accord with the POE.

In \S 2 we present the problems concerned with the energy carried by 
the radiation and the non-existence of the radiation reaction force in certain cases.      
   In \S 3 we present a freely falling system of reference as the preferred 
system to work in,   
 and in \S 4 we calculate the energy carried by the  radiation from the 
 supported charge in a GF,  
 using the work done to overcome the stress 
 force of the field.    We conclude in  \S 5.   

\bg
  
\no {\bf 2. The Problem  } 
\bg

Treating radiation as a relative phenomenon leads to contradictions, because 
radiation transffers energy from one  system to another.  If the energy carried 
by the radiation is absorbed in some system and causes   
  there a certain change, like 
an excitation of a higher energy level, this absorption must be observed by any 
observer, even if he does not have the means to observe directly the flow of 
the energy.  If a static observer observes radiation from a freely falling 
charge, he also must  be able to identify the source of the energy for this 
radiation.   An observer falling freely in parallel to the charge, must observe 
this source of energy, even if he cannot observe directly the radiation that 
carries the energy.  
Similar contradictions arise for the case of a 
 charge supported at rest in a GF,   
 where a static observer does not observe the radiation, and a freely 
falling observer does.  We find that    
 treating radiation as a relative phenomenon leads to contradictions 
concerning  both the source of the energy carried by the radiation, and the 
phenomena that may be caused in absorbing the radiation. The  emmitance of 
radiation is a physical event that cannot be transformed away 
by a coordinate transformation (see [3]).  

There is another difficulty with the common approach - it is generally believed 
that when radiation is created by an accelration, a radiation reaction force is 
created, which contradicts the force that creates the acceleration. The work 
done  by the external force to overcome the reaction force, is considered as 
the source of the  energy carried    
 by the radiation.   However, when the velocity of the charge is low 
$(v\ll c)$, the radiation is emitted mainly 
 in a plane which is perpendicular to 
the direction of motion ([4] pp. 663 and [7]). No momentum is imparted 
to the accelerated charge by the radiation, and no radiation reaction 
  force exists [6].  The source    
 of the energy carried by the radiation should be looked for elsewhere. 

\bg
  
\no {\bf 3. A Freely Falling System of Reference   } 
\bg

According to Jackson [4], a radiation exists whenever an electric charge is 
accelerated.  However, a question should be raised to what system of reference 
this acceleration is related. Without stating it explicitly, Jackson refers to 
 an inertial system of reference.   Ordinarily, when general relativity is 
 considered, the inertial system of reference should be replaced by a freely 
 falling system of reference, characterized by a set of geodesics that covers 
 this system.  The ``absolute acceleration" of a charge supported at rest 
 in a gravitational field does not vanish, where absolute acceleration is the 
covariant time derivative of the four velocity of the charged particle.  A 
 general relativistic criterion for the existence of radiation, is the 
 non-vanishing of the absolute acceleration.    
 A regular acceleration  is related to the system of geodesics that 
 covers the local space.  The preferred 
 system of reference to work in is the system characterized by local 
 geodesics, and freely falling objects - particles and fields - follow these 
 geodesics.  The electric field of a charge is an independent physical entity.  
 Once it is 
 induced on space, its behaviour is determined by the properties of space.  
 When the charge is accelerated by an external (non-gravitational) force, the 
electric field of the charge is not accelerated, and a relative acceleration 
exists between the charge and its field. As was shown by Fulton and  
 Rohrlich [6], the electric field of the charge is  curved.  
 There is a stress force between the charge and its curved field, 
and, as shown in [5],  this force gives rise to radiation. 

A neutral particle and a similar charged particle will fall with the same 
acceleration.   It was shown  
 that the key feature for the creation of radiation is not the relative 
 acceleration between the charge and the observer, but rather the relative 
 acceleration between the charge and its own electric field.  
 
A freely falling charge in a uniform GF follows a geodetic line in this system, 
 and it is not subject to any external force.  The electric field of the charge 
 follows similar geodesics.   The charge and its field both are located in the 
 same frame of reference, and in that frame their relative situation is similar 
 to the one existing between a static charge and its field in a free space. 
 No relative acceleration exists between the charge and its electric field, and 
 we conclude that a freely falling charge does not radiate.

 The creation of radiation by a uniformly accelrated charge was 
 analyzed ([5],[7]),     
 and it was shown that the electric field of the accelrated charge is curved, 
 and there exists a stress force between  the charge and its (curved) field.  
 The stress force $F_s$, is given by: $F_s=E^2/4\pi R_c$, where $R_c$ is the
 radius of  curvature, whose value close to the point charge is:      
    $R_c=c^2/(a \sin \theta)$,  where  $a$  is the acceleration,  
  and $\theta$ is the angle between the direction of the acceleration 
 and the initial direction of the field.    
 By calculating the stress force and the work performed to overcome this   
 force, it is shown that for a uniformly accelerated charge and for  
 very low velocities, the power supplied by the accelerating (external) 
force to overcome the the stress force, equals the power radiated by 
the accelerated charge according to Larmor formula  [5].    
 It is concluded that the  work done in overcoming the stress force 
 is the source of the energy carried by the 
 radiation, and this work is done by the external force that imparts the 
 acceleration to the charge, in addition to the work it does in creating  the 
 kinetic energy of the charge.  
 
\bg
  
\no {\bf 4. A Charge Supported in a Homogenous Gravitaional Field } 

\bg

 The electric field of a charge  supported at rest in the lab 
 against GF  seems static,  but it is not.    
 The  electric field, which is an independent physical entity,  
  is not supported with the charge, and it  
  falls freely in the gravitaional field.  There is a relative acceleration 
 between the charge and its electric field, the field is curved (both in the 
 lab system and in the freely falling system), and a stress 
 force exists between the charge and its field. The 
 (freely falling) electric field     
  follows the system of refernec characterized by the geodesics.     
 To calculate the fields of the supported charge in the freely falling
 geodetic  system, we adopt the results given by Rohrlich [8].   Let us 
 assign primes to the variables calculated in the freely falling 
 system, $S'$.  


  According to Rohrlich, the field equations   
  of the supported charge, in $S'$  are:

$$ E_\rho' = {8 e \alpha^2 \rho' z'\over  \xi'^3}  \eqno(1) $$ 

$$ E_z' = {-4 e\alpha^2 \over \xi'^3} [z_p'^2 + \rho'^2 - z'^2]   
   \eqno(2)   $$ 

$$ B_\phi' = {8 e \alpha^2 \rho' ct' \over  \xi'^3}  \eqno(3) $$ 

 $$ E_\phi' = B_\rho' = B_z' = 0   \eqno(4)  $$ 
 
\no where 

$$ \xi'^2 =  [z_p'^2 - \rho'^2 - z'^2]^2 + (2\alpha \rho')^2  \eqno(5) $$     
\no where we used  for the particle location: $ z_p'^2 = \alpha^2 + (ct')^2 $, 
and $\alpha = c^2/g $ is the particle location at $t' = 0$.  
 Certainly, the Poynting vector  does not vanish in this system.  

Using transformations 
 given by Rohrlich [8] we can calculate the electromagnetic fields in the 
 lab system. It follows (as can be expected), that the magnetic filed 
 vanishes in this system, and the Poynting vector vanishes as well.  This 
led Rohrlich  to conclude that a charge supported at rest 
 in a gravitational field does not radiate.  However,     
we know that a Poynting vector is not an invariant [9], 
and we demand that 
the existence of radiation must be represented by a non-vanishing 
Poynting vector in the frame of reference characterized by the 
local geodesics, $S'$, and {\it in this system the Poynting vector 
 does not vanish}. 

   The situation is not static, 
 and  the electric field exists in a steady state.  The pattern of 
the electric field remains constant, but the field itself does 
not.  As we emphasized earlier, the electric field is a property 
of the space on which it was induced, and its behaviour is 
determined by this space.  The electric field  is detached 
from the supported charge, and it is not supported against 
gravity as the charge is.  Hence the electric field falls in a 
free fall, and it has an acceleration $g$ relative to the 
 supported charge.   
 In the freely falling system, which also has an acceleration $g$ 
relative to the supported charge,  the charge is accelerated 
 upward with an acceleration  $g$.  


 It was also shown 
 by Rohrlich [8], that in the system characterized by the geodesics, a magnetic 
 field does exist, and it comes out that the Poynting vector does not vanish. 
We conclude   
 that a charge supported at rest in a gravitational field does radiate. 
 In Figure 1 we present the curved elcetric field lines calculated for an 
electric charge supported at rest in a uniform homogenous GF, 
characterized by an acceleration  $g$.  The field is similar to the 
 one calculated by  Singal [10], for a uniformly accelerated charge. 

\begin{figure}
\centering\epsfig{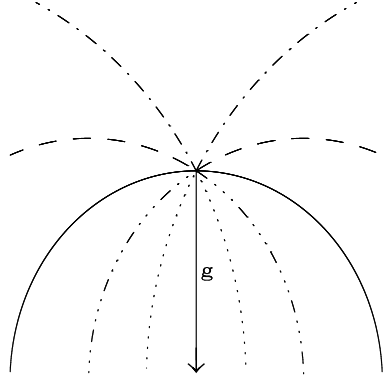} 
\vskip 2cm
\caption{A curved electric field of a charge supported in a uniform 
homogenous gravitational field.} 
\end{figure} 
\bg

 The curved electric field gives rise to a stress force, and we 
 calculate the work done in overcoming this force in a way similar 
 to that used in  [5] for the uniformly accelerated charge, 
 where the calculations are carried now in the (flat) freely 
 falling system of reference.   

For the sake of convenience we ommit now the primes.   
 We shall sum over the stress force of the field, $f_s$, and 
calculate the work done against this force.   
In order to sum over $f_s$,   we have to 
integrate over a sphere whose center is located on the charge.  
Naturally, such an integration involves a divergence (at the center). 
To avoid such a divergence, we take as the lower limit of the 
integration a small distance from the center, $r = c \Delta t$, (where 
$\Delta t$ is infinitsimal), and later we demand that 
$\Delta t  \rightarrow 0$.  
  We calculate the work done by the stress 
force in the volume defined by  $c \Delta t < r < r_{up}$, 
where  $c^2/g \gg r_{up} \gg c \Delta t$.   
These calculations are performed in the geodetic system (the 
  system of reference defined  
 by the geodesics), which momentarily coincides with the frame of 
reference of the  charge at the charge location, 
at time $t = 0$.   

 The force per unit volume  due to the electric stress is 
$f_s = E^2/(4 \pi R_c)$, where $E$  is the electric field, and 
$R_c$ is the radius of curvature of the   
field lines.
 The radius of curvature is: $R_c \simeq  c^2 / (g \sin \theta)$,
where $\theta$ is the angle between the initial direction of the 
 electric field line and the direction of the acceleration $g$  of 
the charge, as seen in the geodetic system.     
 The force per unit volume due  to   the electric stress is

$$   f_s (r) = {{E^2}(r)\over{4 \pi R_c}} = 
 {{g \sin \theta}\over{c^2}} {{e^2}\over{4 \pi r^4}} , \eqno(6)  $$   

\no where in the second equality we have substituted for the electric field  
 $E=e/r^2$, which is a good approximation in weak graviatational fields [8]. 
 The stress force is perpendicular to the direction of the field lines, 
so that the component of the stress force along the acceleration $g$ is
$ - f_s(r) \sin \phi$, where  $\phi$ is the angle between the 
local field line and the acceleration.   For very short intervals (where 
the direction of the  field lines did not change much 
 from their original  direction) $\phi \sim \theta$, 
and we can write: 
$ - f_s(r) \sin \phi \simeq - f_s(r) \sin \theta  = 
  {-g \sin^2 \theta \over c^2} {e^2 \over 4 \pi r^4}$.   
The dependence of this force on $\theta$ is similar to the dependence of the radiation 
distribution of an  accelerated charge at zero velocity on $\theta$.      
Integration of this force  over a spherical shell extending from $r=c \Delta t$ 
to $r_ {\rm up}$ (where $c^2 / g \gg r_{\rm up} \gg c \Delta t$),
yields the total force due to the stress 

$$  F_s(t) = 2 \pi  \int _{c \Delta t}^{r_{\rm up}} r^2 dr 
\int_0^{\pi} \sin \theta d \theta
[- f_s(r) \sin \theta ]
= - {{2}\over{3}} {{g}\over{c^2}} {{e^2}\over{c \Delta t}}
\left(1-{{c \Delta t}\over{r_{\rm up}}} \right) . \eqno(7)  $$
 
 Clearly the second term in the parenthesis can be neglected.
The power created in overcoming the  
electric stress force is: 

     $$P_s= - F_s  v = - F_s g \Delta t , \eqno(8) $$  
 
\no where we substituted  $v= g \Delta t$, and  $v$  is the charge velocity 
 in the geodetic system,  at time $t = \Delta t$.   
Substituting for  $F_s$   we obtain (at the limit $\Delta t \rightarrow 0$):  

$$ P_s(t) = {{2}\over{3}} {{g^2 e^2}\over{c^3}}  \eqno(9)   $$  

 \no which  is equivalent to the power radiated by an accelerated charged 
 particle (Larmor formula), where the acceleration is 
replaced by  $g$.   Thus we find that the work done 
 against the stress force, supplies the energy carried by the 
radiation.

 Who is  performing this work or, 
 what is the source of the energy of the  radiation? 
 
     The charge is supported by a solid object, which is
    static in  the GF.  This
     solid objet must be rigidly connected to the source of the
 GF.  Otherwise, it will fall in the GF, together with the "supported"
     charge.   This means that actually, the supporting object  is
     part of the object that creates the GF.

     As we already mentioned, the charge is static and no work
     is done by the GF that acts on the charge.  However, the
     electric field of the charge is not static, and it falls in a
     free fall in the GF.  If there was no interaction between the
     electric field and the charge that induced the field, the
      field would have follow a geodetic line and no work would have
       been needed to keep it  following the geodetic line. But the
     field is curved, and a stress force is implied.  The interaction
     between the curved field and the supported charge creates a
     force that contradicts the free fall.   In order to overcome this
force and cause the electric field to follow the geodetic lines, a 
     work should be done on the electric field, and this work is
     done  by the GF.  This work is the source of the energy 
 carried by the radiation.  It comes out     
     that the energy carried away by the radiation is supplied by the
     GF, that loses this energy.

\bg
\no{\bf  5.  Conclusions }
\bg
 
 It is found that the ``naive" conclusion from the principle of equivalence - 
that a freely falling  charge does not radiate, and a charge supported 
 at rest in a gravitational field does radiate - is a correct conclusion, 
 and one should look for rdiation whenever a relative 
  acceleration exists between 
an electric charge and its electric field.   The electric field which falls  
 freely in the gravitational field is accelerated relative to the static 
charge.   The field is curved, and the work done in overcoming the stress 
 force created in the curved field, is the source of the energy carried by 
 the radiation.  This work is done by the gravitational field on the 
 electric field, and the energy carried by the radiation is created in 
the expence of the gravitational energy of the system.      

Motz [11] suggested that  the huge radiation emerging 
 from quasars may be created by charges located in the strong 
 gravitational fields close to the surface of the quasars.  Although 
 the current expalnation for this phenomenon is different, radiation from 
 charges located in strong gravitational fields can still play a role 
 in certain cosmological phenomena.  
 
    We conclude that we find both the mechanism that creates the radiation
     emitted by a charge supported in a GF, and  the source of the
     energy carried by this  radiation.

\bg

{\bf {ACKNOWLEDGEMENT}}

\bg

  We acknowledge useful discussions 
on this topic with Amos Ori from the Technion.

\vfil

\eject

 \bg 

 \no {\bf references:}   
 
 \bg

 \no [1]  Rohrlich, F. 1965, in {\it Classical Charged Particles}, 
  Addison-Wesley Pub. Co., MA.   
   
 \no [2]  Boulware, D. G. 1980, Annals of Physics, 124, 169.

  \no [3]  Matsas, G.E.A., 1994,  
  Gen. Rel. Grav.,  26, 1165.

 \no [4]  Jackson, J. D. 1975, {\it Classical Electrodynamics}, 
Second Edition, John Wiley \& Sons (New York). 

 \no [5] Harpaz, A., Soker, N., 1998,  
    Gen. Rel. Grav., 30, 1217.
   
 \no [6] Fulton, T., Rohrlich, F., 1960, Annals of Physics, 9, 499.

 \no [7] Harpaz, A., Soker, N., 1999, in {\it Proceedings of the $4^{th}$  
 Freedmann Int. Seminar on Gravitation and Cosmology}, edts., 
Yu.N. Gnedin, A.A. Gribs, V.M. Mostepanenko, W. Rodrigues Jr., 
UNICAMP (Br), and Friedmann Lab. Pub. (St. Petersburgh).  
 
 \no [8] Rohrlich, F., 1963, 
  Annals of Physics, 22, 169.

\no [9] Parrot, S., 1997, 
 paper 9303025, archive
 gr-qc@xxx.lanl.gov.

 \no [10] Singal, A.K., 1997,  
    Gen. Rel. Grav., 29, 1371.

 \no [11]  Motz, L.A.   1972, Nuovo Cimento, 9B, 77.   

\end{document}